\documentclass[a4paper,11pt]{article}
\usepackage{pos}
\usepackage{microtype}
\usepackage{chemformula}
\usepackage[separate-uncertainty=true]{siunitx}
\usepackage{todonotes}

\title{Commissioning of a mobile neutron spectrometer for LNGS}

\author*[a,b,c]{Francesco Pompa}
\author[c]{Klaus Eitel}
\author[a,b]{Alfredo Davide Ferella}
\author[d]{Felix Kratzmeier}
\author[d,e]{Melih Solmaz}
\author[c,d]{Kathrin Valerius}

\affiliation[a]{Department of Physics and Chemistry, University of L’Aquila, 67100 L’Aquila, Italy}
\affiliation[b]{INFN, Laboratori Nazionali del Gran Sasso, I-67100 Assergi, Italy}
\affiliation[c]{Institute for Astroparticle Physics, Karlsruhe Institute of Technology, 76021 Karlsruhe, Germany}
\affiliation[d]{Institute of Experimental Particle Physics, Karlsruhe Institute of Technology, 76021 Karlsruhe, Germany}
\affiliation[e]{Kirchhoff Institute for Physics, Heidelberg University, 69120 Heidelberg, Germany}

\emailAdd{francesco.pompa@graduate.univaq.it}

\abstract{Environmental neutrons are a source of background for rare event searches in underground laboratories. Since the majority of the neutron background comes from the cavern walls due to the intrinsic radioactivity of concrete and rock, the flux is known to be time and location dependent. Therefore, a precise knowledge of the spectrum and of the total flux is needed to devise shielding and veto mechanisms for rare event searches. Here ALMOND (An LNGS Mobile Neutron Detector) is presented. It is a mobile neutron spectrometer, based on capture-gated spectroscopy and comprised of an array of plastic scintillator bars wrapped with gadolinium foils. The detector has been calibrated with Americium-Beryllium source at Karlsruhe Institute of Technology and with an Americium-Boron source and a D-D generator at ENEA Frascati. The results of the neutron calibration with the time of flight method and the D-D generator are shown here, alongside the first results on capture time profile. Moreover, the first results from the neutron background run in Hall A at LNGS are presented.}

\FullConference{The European Physical Society Conference on High Energy Physics (EPS-HEP2025) \\
6-11 July, 2025\\
Marseille, France
}

\begin{document}
\maketitle

\section{LNGS neutron background}

LNGS hosts a variety of experiments dedicated to rare event searches. An accurate knowledge of the neutron background is required to design neutron veto and shielding mechanisms. In underground laboratories, the neutron background \cite{wulandari_neutron_2004} originates mainly from spontaneous fission of \ch{^{238}U} and ($\alpha$,n) reactions
on light nuclei initiated by $\alpha$-decays of the uranium
and thorium decay chains.
The neutron background is known to be location and time dependent because of the different concentrations of radioactive nuclei in the concrete and rock and seasonal variations in the water levels. 

ALMOND (An LNGS Mobile Neutron Detector) \cite{solmaz_design_2023} is designed to reduce the systematics in the measurement of the neutron background by measuring the neutron flux in different locations with the same technique and by taking advantage of capture-gated spectroscopy.

\section{Detector Layout}

\begin{figure}[htbp]
    \centering
    \includegraphics[height=0.35\linewidth]{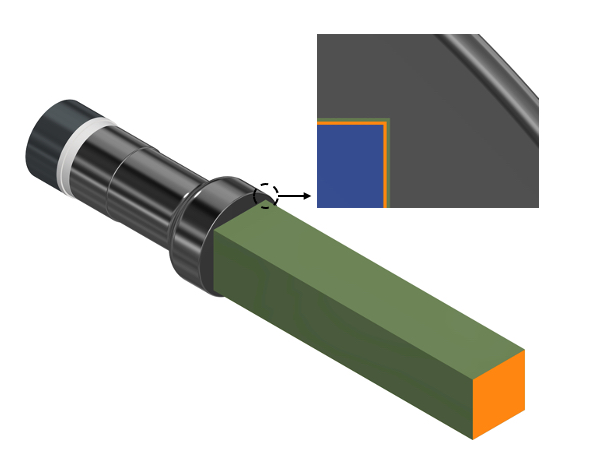} \quad
    \includegraphics[height = 0.35 \textwidth]{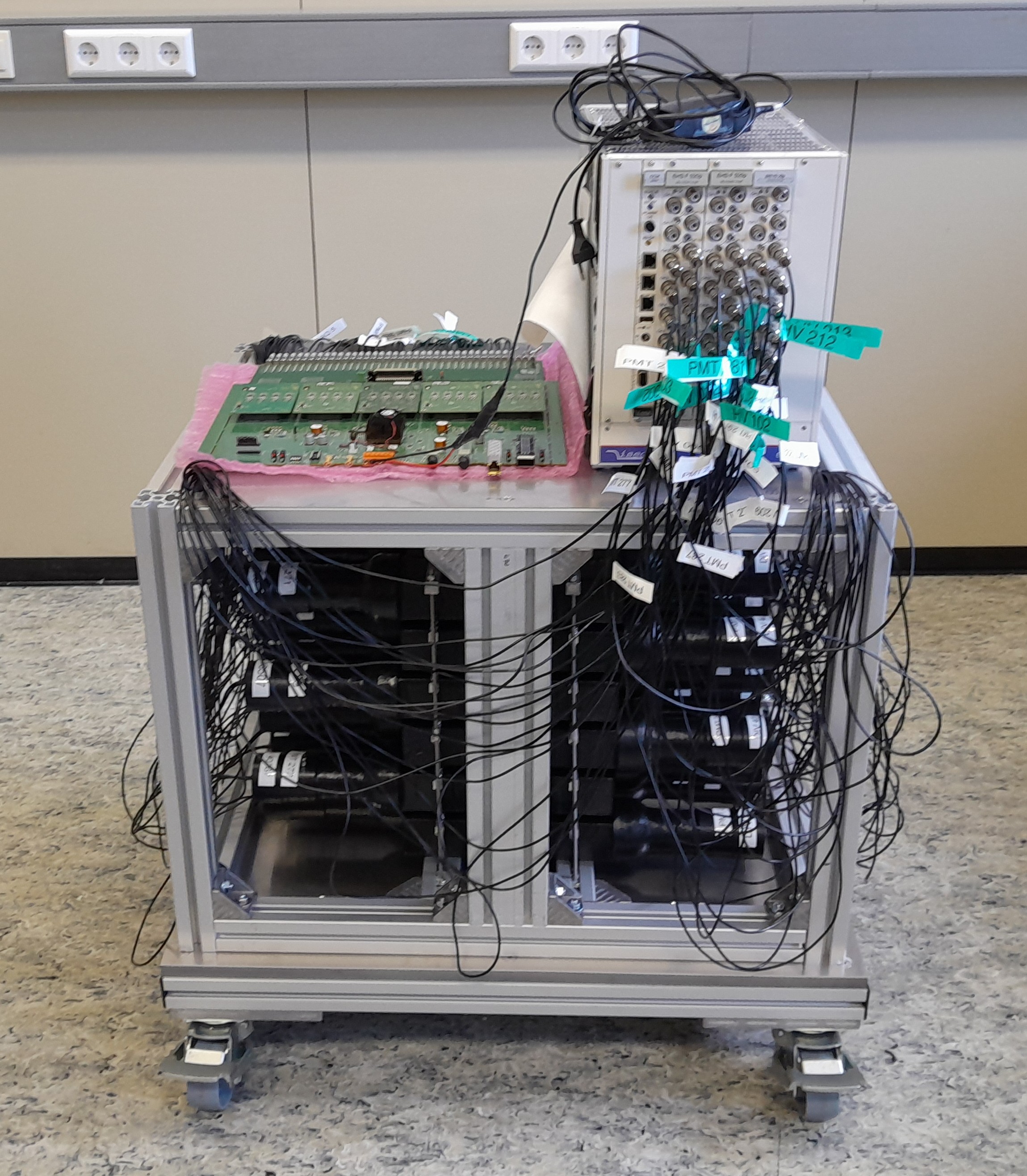}
    \caption{Left: scheme of a single ALMOND module with cross sectional view of a corner near the PMT. In blue, the EJ-200 scintillator, in orange, the reflector foil, in green, the \SI{100}{\micro m} gadolinium foil and in black the low background ET 9302B PMT. Right: the detector in the calibration setup. A Bosch profile structure houses an array of $6 \times 6$ modules, DAQ board and HV supply on top, here without Pb sheets on side walls.}
    \label{fig:ALMOND_detector}
\end{figure}

ALMOND (Figure~\ref{fig:ALMOND_detector}) is made of 36 EJ-200 plastic scintillators with dimensions $(5 \times 5 \times 25$)\,cm$^3$ wrapped in reflector foil to increase light collection efficiency and gadolinium foils
for neutron capture. Each scintillator bar is coupled to
a 3-inch photomultiplier tube (PMT) to detect scintillation light.
The capture process produces a gamma cascade with an energy well above the \SI{2615}{keV} end point of the environmental gamma background. The detector is equipped with a \SI{16}{mm} thick lead shield to reduce the impact of the gamma background, which, in the case of pile-up,
could be misidentified as the gadolinium neutron capture signal.

The DAQ is designed to trigger on the high-energy gamma cascade from gadolinium capture using a threshold set above natural radioactivity, typically \SI{3}{MeV_{\text{ee}}}. Then, all pulses within a custom interval up to $\pm \SI{100}{\micro s}$ are recorded based on a secondary lower threshold. All pulse traces are sampled with a sampling frequency of \SI{62.5}{MS/s}.

\section{Americium-Beryllium calibrations}

\begin{figure}[htbp]
    \centering
    \includegraphics[width=0.8\linewidth,trim={0 0.7cm 1cm 1cm},clip]{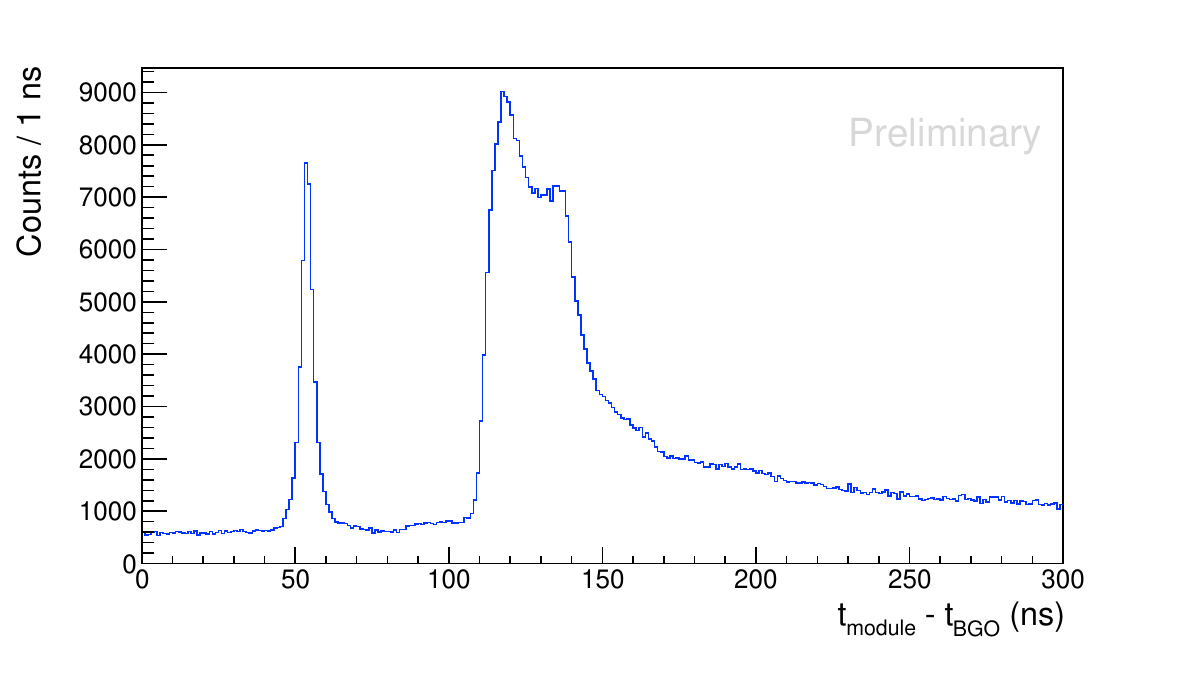} 
    \caption{Time difference between the module and the BGO in the time-of-flight measurement. The peak on the left is due to coincidence between two gamma rays, the structure on the right is given by delayed coincidence between the neutron and the \SI{4.4}{MeV} gamma.}
    \label{fig:deltaT_TOF}
\end{figure}

Before integrating all modules into the detector array, a single module was calibrated at Karlsruhe Institute of Technology using the time-of-flight method. An Americium–Beryllium (AmBe) source was placed near a bismuth germanate (\ch{Bi4Ge3O12}, hereafter BGO) detector and at a distance of \SI{2}{m} from the scintillator module. The AmBe source emits a neutron and a gamma ray simultaneously about 60\% of the time.
By detecting the gamma ray with the BGO and measuring the delay to the neutron detection in the scintillator, it is possible to determine both the neutron kinetic energy and its light yield in the scintillator. The fit of the data with Birks’ law gives a Birks' constant $kB$ compatible with the results previously obtained in Laplace et al. \cite{laplace_low_2020} and Langford et al. \cite{langford_development_2016}. The distribution of the time difference between the ALMOND module and the BGO is shown in Figure~\ref{fig:deltaT_TOF}. The gamma and neutron populations can be distinguished as the two structures in the figure.

After full construction of the detector, another AmBe measurement was performed. The BGO detector was used, as before, to tag neutrons, but time-of-flight could not provide an independent estimate of the neutron energy due to limits in time resolution. Instead, the neutron capture time profile was characterized without relying on BGO tagging. Pulses were classified in clusters by imposing a minimum separation of 5 samples between them, corresponding to \SI{90}{ns}. Only events containing a single cluster were selected to restrict the population to events that can be correctly reconstructed by the DAQ. In fact, when the number of pulses in coincidence is too high, there is the possibility of loss of data. After background subtraction by removing accidental coincidences occurring after the main trigger, the capture time profile can be determined. The results will be compared with a similar Americium–Boron (AmB) calibration in the following section.


\section{Calibrations at ENEA Frascati}

After extensive testing for proper functioning, ALMOND was calibrated at the Frascati Neutron Generator (FNG) \cite{pietropaolo_frascati_2018} to make use of the D-D generator and the AmB source. The D-D generator emits monoenergetic neutrons at a nominal energy of 2.4 MeV. However, the energy depends on the angle of emission. Since the detector was placed at an angle greater than \SI{150}{\degree}, the neutron kinetic energy is estimated to be \SI{2.2}{MeV} \cite{chichester_measurement_2011}.

\begin{figure}[htbp]
    \centering
    \includegraphics[width=0.55\linewidth]{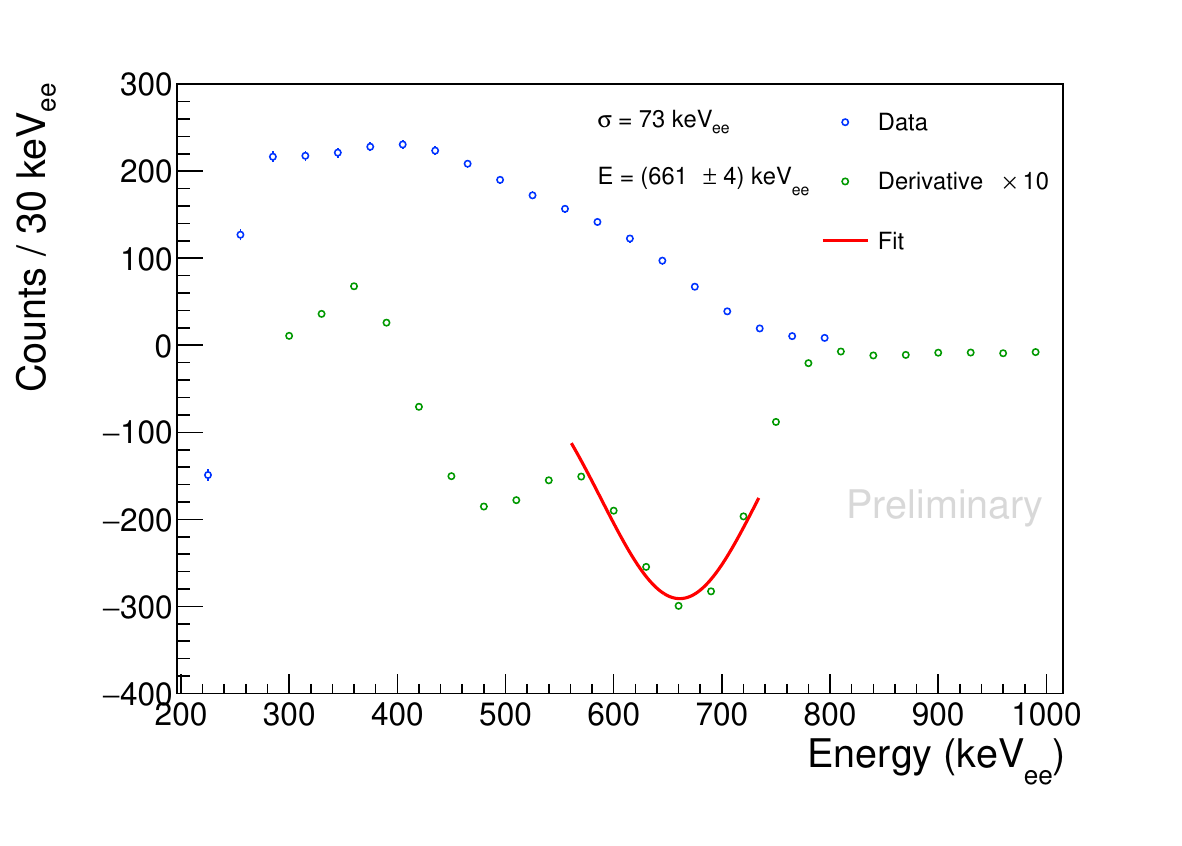} \\
    \includegraphics[width=0.5 \linewidth,trim={0cm 1.2cm 2cm 1cm},clip]{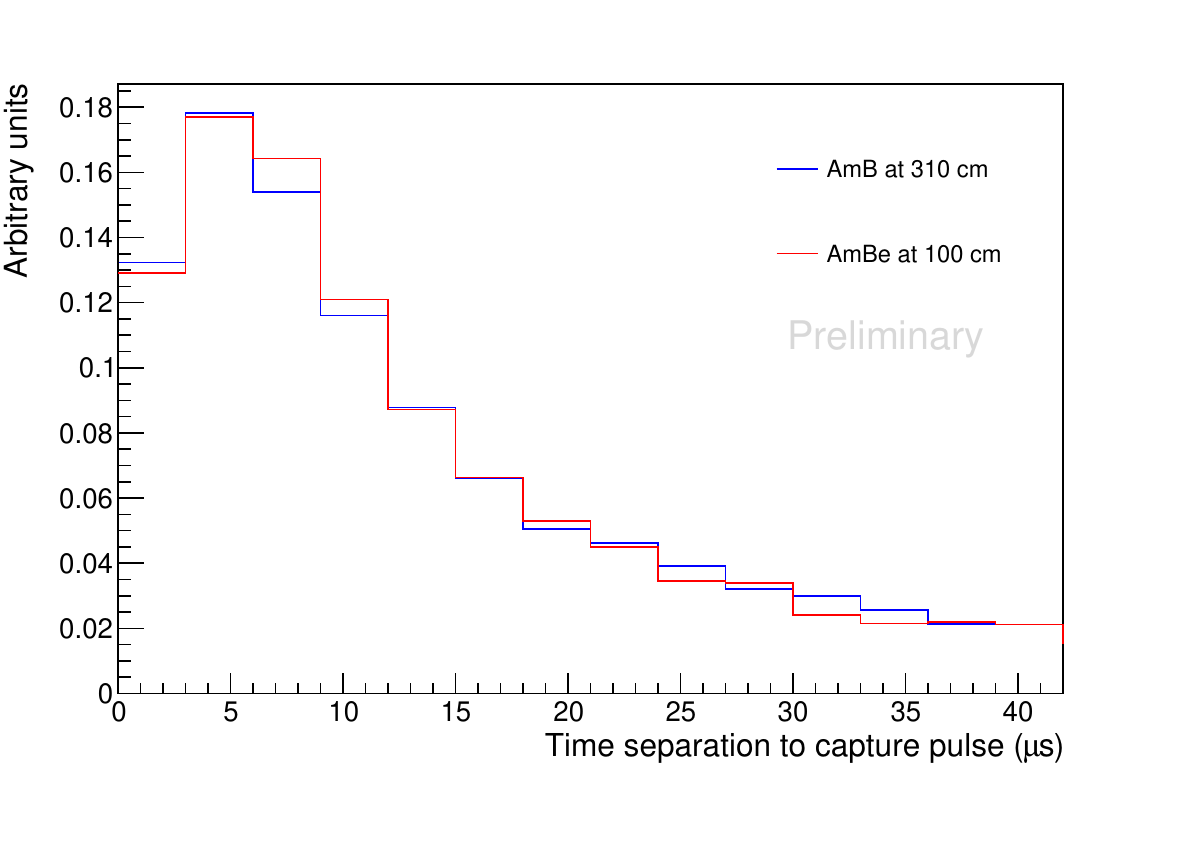}
    \caption{Top: fit of the proton recoil edge after background subtraction leading to an energy deposit of \SI{654 \pm 4}{keV_{\rm ee}}
    produced by 2.2 MeV D-D neutrons. Bottom: comparison of the capture time profiles between the AmBe calibration at KIT and the AmB calibration at FNG.}
    \label{fig:D-DfitAmBtimeProfile}
\end{figure}

For the energy calibration, the energies of all pulses recorded before the trigger are used, and duplicate energies due to close triggers are counted only once. From time-of-flight measurements, the recoil signal from D-D neutrons is expected to extend up to around \SI{700}{keV_{\text{ee}}}. Beyond this region, the background can be modeled empirically as a power law:
\begin{equation}
f(E) = A E^{-\gamma},
\end{equation}
with $\gamma \sim 1.2$. After applying background subtraction, the derivative of the spectrum is fitted to determine the proton recoil edge \cite{kornilov_total_2009}. The fit is shown in the top panel of Figure~\ref{fig:D-DfitAmBtimeProfile}. The capture time profile
(bottom panel of Figure~\ref{fig:D-DfitAmBtimeProfile})
is derived using the same method described in the previous section and is consistent with the previous AmBe calibration.

\section{Commissioning at LNGS and outlook}
\begin{figure}[htbp]
    \centering
    \includegraphics[width=0.8\linewidth]{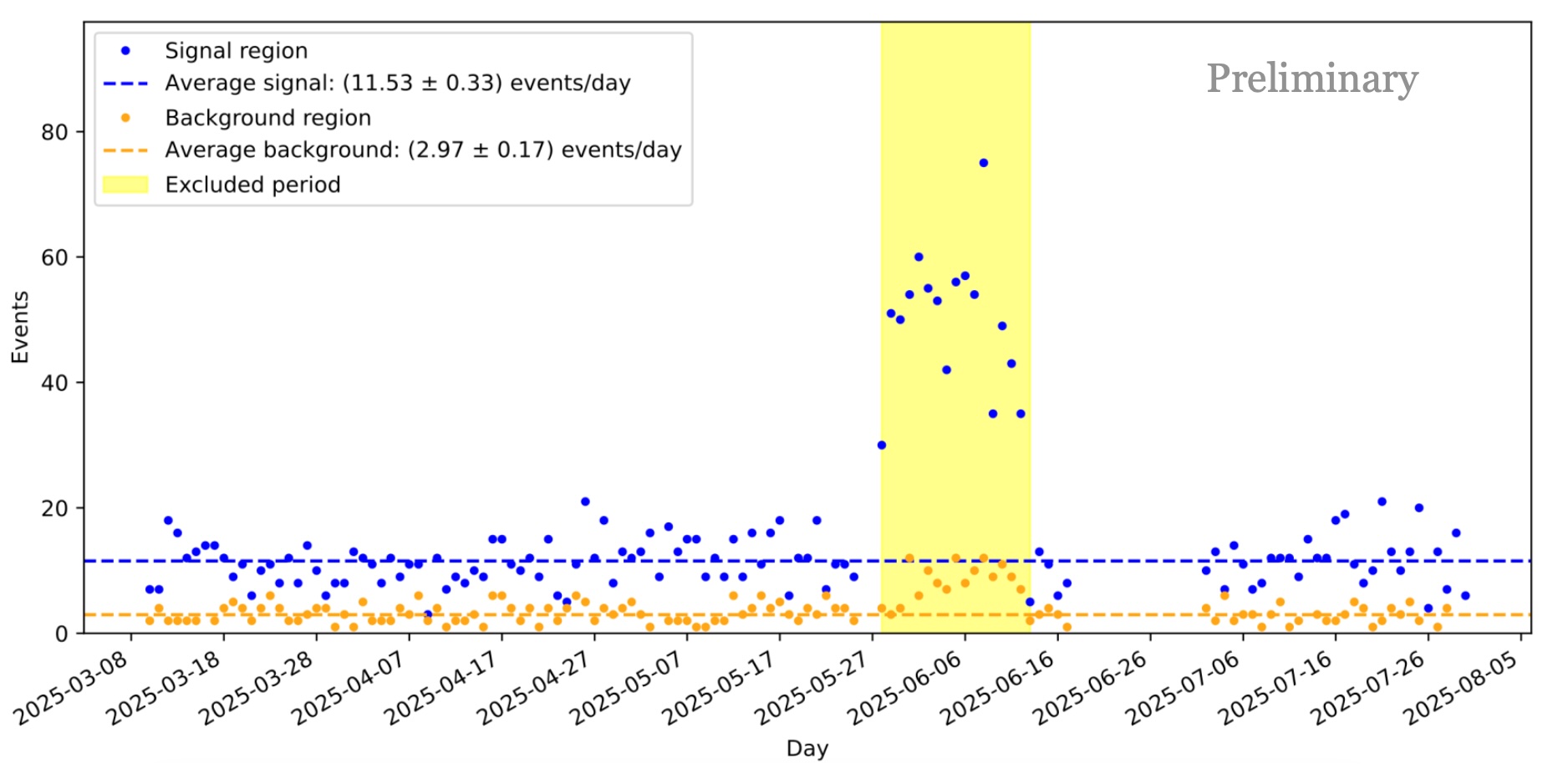}
    \caption{Neutron candidates and accidental coincidences per day for the neutron background run in Hall A at LNGS. The region in yellow was excluded due to the presence of a neutron source in proximity of the detector.}
    \label{fig:neutronBackgroundHallA}
\end{figure}

After the calibration at the FNG, ALMOND was commissioned at the LNGS in February 2025. Once it was verified that the rates of accidental coincidences and intrinsic neutron background were sufficiently low, a neutron background run was initiated in Hall A. The rates of neutron candidates and accidental coincidences within a time window of \SI{40}{\micro s} around the main trigger, with a capture threshold of \SI{3}{MeV_{\rm ee}} and a proton recoil threshold greater than \SI{20}{keV_{\rm ee}} are shown in Figure~\ref{fig:neutronBackgroundHallA}. The detector meets the design requirements and is sensitive to the ambient neutron flux at LNGS, which is on the order of $10^{-6} \, \si{n/cm^2/s}$.

In August 2025, a new run was started in Hall C. ALMOND will be deployed in other areas of the LNGS to further improve the characterization of the neutron background. It will also serve as an important addition to the research infrastructure at the laboratory.

\acknowledgments{
This work has received financial support from the German Federal Ministry of Research, Technology and Space (BMFTR) under the grant number 05A21VK1 and the Italian National Institute for Nuclear Physics (INFN). We thank our colleagues at ENEA Frascati for providing access and technical assistance during the ALMOND calibration at the Frascati Neutron Generator. 
}

\clearpage



\bibliographystyle{JHEP}
\bibliography{proceedings_references}

\end{document}